\documentclass[twocolumn,preprintnumbers,amsmath,amssymb]{revtex4}

\usepackage{graphicx}
\usepackage{dcolumn}
\usepackage{bm}
\usepackage{epsfig}
\usepackage{amssymb}
\begin{document}

\preprint{}

\title{Comments on ``Final Analysis and Results of the Phase II \\ SIMPLE Dark Matter Search''}

\author{J.I. 
Collar$^{a}$ 
}
\address{ 
$^{a}$Enrico Fermi Institute, Kavli Institute for Cosmological Physics and Department of Physics, University of Chicago, Chicago, IL 60637\\
}

\maketitle

The SIMPLE collaboration has recently claimed \cite{phaseIIb} improved limits on WIMP-nucleus couplings. The limited shelf-life of SIMPLE detectors and apparent deficiencies in the acoustic background discrimination method employed severely affect the credibility of these claims. 

SIMPLE and PICASSO \cite{picasso} use dispersions of superheated droplets  in a water-based gel (SDDs) to search for WIMPs, putative dark matter candidates. Their detector design is however very different. PICASSO employs hermetically enclosed SDD modules with no inert volume above the active matrix, and periodic pressure cycling able to re-compress existing bubbles and to heal any incipient gel fractures. This leads to a demonstrated detector shelf-life sufficient for long exposures. This design is similar to that implemented in commercial SDDs, neutron "bubble dosimeters". In contrast to this, SIMPLE modules contain a considerable volume of inert liquid (glycerol) and gas (compressed N$_{2}$) above the active gel, and cannot be recompressed due to the fragility of their glass containers. This produces a rapid aging of the detector material, leading to gel fractures and a diffusion of the superheated liquid into inert volumes and fracture voids, increasing the pressure inside the module and ultimately leading to gas leaks occasionally able to produce characteristic acoustic signals. While fractures, pressure increase and gas leaks are acknowledged by the SIMPLE collaboration \cite{phaseIIb, phaseIIa}, their origin in module deterioration is not emphasized in these recent publications. This is however treated in earlier communications \cite{njp,aging}. This aging, aggravated by increased operating temperature, leads to a noticeable depletion of the active superheated liquid, and a subsequent long-term decrease in the response of SIMPLE modules to neutron and alpha sources \cite{puibasset}. 

Early SIMPLE results \cite{oldprl} were limited to short initial exposures (20 d, of which only 8 d were above 9$^{\circ}$C) because of these known limitations. The spontaneous nucleation rate above 9$^{\circ}$C was observed to decrease by an order of magnitude after $\sim$45 d of exposure during runs following those in \cite{oldprl} (see discussion around Fig.\ 5.13 in \cite{puibasset}). This is in contrast to the most recent SIMPLE results, which implicitly assume perfect detector stability for periods of 90-100 d at 9$^{\circ}$C. This is done without any neutron calibrations during or following runs to demonstrate a sufficiently constant detector response. More than a decade after \cite{oldprl}, SIMPLE modules feature only marginal improvements to the original module design, and are still not able to contain target mass diffusion or gas leakage \cite{phaseIIb, phaseIIa,nimelectronics}. The depletion of the active volume in SIMPLE modules after more than three months of exposure at 9$^{\circ}$C should be readily visible as an increased transparency and change in gel coloration. 

Other important concerns can be expressed. For instance, contrary to the claims in \cite{phaseIIa}, events identified as recoil-like during physics runs via their acoustic signature are markedly different from those induced during neutron calibrations (see Fig.\ 1 in \cite{phaseIIa}), pointing at a different origin. This is most probably found in environmental acoustic noise, for which no unambiguous discrimination criteria can be discerned in Fig.\ 3a of \cite{nimelectronics}, or in \cite{tom}. As an example of the probable misidentification of these events as recoil-like, the single event claimed to pass cuts in \cite{phaseIIb} is alleged to have a neutron origin based on a double acoustic signature separated by 30 ms. However, the straggling of fast neutrons prior to thermalization in hydrogenated materials is never longer than few tens of $\mu$s. Secondarily, the WIMP limits derived in \cite{phaseIIb,phaseIIa} involve the subtraction of a simulated environmental neutron contribution, based solely on Monte Carlo simulations and not on an actual measurement of the neutron flux and energy spectrum at the experimental site. Such predictions are uncertain: the SIMPLE collaboration is isolated from all other dark matter detection efforts in this daring approach to limit extraction. 

The SIMPLE collaboration is invited to demonstrate a sufficient detector longevity via dedicated neutron calibrations. Based on previous experience with these SDDs, a very significant relaxation of the sensitivity claimed in \cite{phaseIIb, phaseIIa} is expected from such tests.

\end{document}